\documentstyle[12pt,titlepage,fleqn]{article}

\newcommand{\be}{\begin{equation}}
\newcommand{\ee}{\end{equation}}
\newcommand{\ba}{\begin{eqnarray}}
\newcommand{\ea}{\end{eqnarray}}

\newcommand{\non}{\nonumber}
\newcommand{\oh}{\frac{1}{2}}
\newcommand{\IIA}{type IIA superstring}

\begin{document}

\begin{titlepage}
\hfill{ hep-th/9604107}
\vspace*{0.1truecm}
\begin{center}

{\large\bf  A SUPERMEMBRANE DESCRIPTION OF STRING-STRING DUALITY}

\vspace{1cm}

{\large Fermin ALDABE\footnote{E-mail: faldabe@phys.ualberta.ca}}

\vspace{1cm}

{\large\em Theoretical Physics Institute,
University of Alberta\\
Edmonton, Alberta, Canada, T6G 2J1}

\vspace{.5in}

\today \\

\vspace{1cm}
{\bf ABSTRACT}\\
\begin{quotation}

\baselineskip=1.5em

We show that the open membrane action on $T^3\times S^1/Z^2$ is
equivalent to the close membrane action on K3.
The main difference
between the two actions is that one generates the KK modes in the
worldvolume action which is the strong coupling limit of \IIA\ while the
other action generates the KK modes in a worldsheet action.
 Thus explaining membrane-string
duality in D=7, which naturally leads to string-string duality in D=6.

\end{quotation}
\end{center}

\end{titlepage}

The spacetime
membrane/string duality in seven dimensions was first presented in \cite{T}.
This duality can be rephrased, after identification of the orbifolded membrane
with the heterotic string \cite{yO, HW}, 
in terms of close/open membrane 
duality.
Our strategy to the worldvolume description of membrane/membrane duality will
be to show that the bosonic sector of 
the open membrane action on $T^3\times S^1/Z_2$ is the same as the
bosonic sector of the  close membrane action on K3 after
both theories are dimensionally reduced.  
The worldvolumes of the close and open membrane are topologically distinct.
However, upon dimensionally reducing each theory, the worldvolumes 
become close strings and are therefore topologically equivalent. 
This will insure that
both theories have the same bosonic massless
spectrum, which provided they have the same spacetime 
supersymmetry insures that both theories have the same massless spectrum.

The action for the bosonic sector of an anomaly free open membrane is \cite{yO}
\ba
S&=&S_M+\int_{\partial M^3}
\{\oh (g_{mn}\eta^{ij}+b_{mn}\epsilon^{ij})\partial_ix^m\partial_jx^n
\non\\
&&+ 
\oh (g_{IJ}\eta^{ij}+b_{IJ}\epsilon^{ij})\partial_ix^I\partial_jx^I
+ \epsilon^{ij}\partial_i x^{J}\partial_j x^{m} A^{J}_{m}(x)\}
\label{S}
\ea
where
\be
S_M=\int_{M^3}  (\sqrt{-g_{mn}\partial_i x^m\partial_j x^n
}+\frac{1}{6}
\epsilon^{ijk}\partial_i x^m\partial_j x^n\partial_k x^p B_{mnp}
),\label{s}
\ee
$g_{mn}$ is the metric on $M^{11}$, $x^m$ are coordinates on $M^{11}$,
and $B_{mnp}$ is an antisymmetric 3-tensor. The worldvolume $M^3$ 
is ${ R}\times S^1\times S^1/Z_2$.
The $16$ left moving bosons $x^{J},\ J=1,...,8$ live only on the boundary
of the open membrane,
which is two copies of ${R}\times S^1$. 
They couple naturally to 
$A^{J}$, 
the $U(1)$ connections.

The double dimensional reduction of the twisted supermembrane on 
$M^{10}\times S^1/Z_2$ of (\ref{S})
was given in \cite{yO}.  The bosonic sector is that of the heterotic string
\ba
S_h&=&\int d^2\sigma
\{\oh (g_{mn}\eta^{ij}+b_{mn}\epsilon^{ij})\partial_ix^m\partial_jx^n
\non\\
&&
+ \oh (g_{IJ}\eta^{ij}+b_{IJ}\epsilon^{ij})\partial_ix^I\partial_jx^I
+\epsilon^{ij}\partial_i x^{I}\partial_n x^{m} A^{(I)}_{m}(x)\}
,\label{Shet}
\ea
and the gauge group indices are now $I=1,...,16$.

We now consider the case in which $M^{10}=T^3\times M^{7}$
where $dim\ H^1(M^{7})=0$.  The worldsheet action
is  a sum of three terms
\ba
S_{het}&=&S_{st}+S_{KK}+S_{mod}\non\\
S_{st}&=&\int d^2\sigma
\oh (g_{mn}\eta^{ij}+b_{mn}\epsilon^{ij})\partial_ix^m\partial_jx^n\non\\
S_{KK}&=&\int d^2\sigma
\epsilon^{ij}\partial_i x^{I}\partial_j x^{m} A^{I}_{m}\non\\
S_{mod}&=&
\int d^2\sigma
 \oh (g_{IJ}\eta^{ij}+b_{IJ}\epsilon^{ij})\partial_ix^J\partial_jx^I
\label{hete}
\ea
The index $I=1,...,22$ labels $22$ gauge fields: 16 come from the
internal dimensions of the heterotic string, and the other 6 gauge fields are
the KK modes of the metric and antisymmetric tensor.  
Sixteen of the $x^I$'s 
are left moving bosons,  the remaining ones are both left-right moving
bosons.  The action $S_{mod}$ has a massless spectrum given by moduli fields
which correspond to deformations of the Narain lattice and therefore
take values on the group manifold
\be
\frac{SO(19,3)}{SO(19)\times SO(3)}.
\ee
Something is very peculiar to action (\ref{hete}).  All the gauge fields
have appeared within a two-dimensional theory, and not a three-dimensional
theory.  This is exactly what  happens in the long wavelength limit of the
open membrane:  the gauge fields are defined in terms of 
fields which live on 10-dimensional boundaries of M-theory.
This does not happen in the long
wavelength limit of the close membrane: there, the gauge fields are defined
in terms of 11-dimensional fields.  We should then expect the gauge fields
of the close membrane to be defined over $M^3$ and not over its boundary which
is absent for the close membrane.

The closed membrane action on $K3\times M^{7}$ is
\be
S'_M=\int_{M^3} d^3\zeta (\sqrt{-g_{mn}\partial_i x^m\partial_j x^n
}+\frac{1}{6}
\epsilon^{ijk}\partial_i x^m\partial_j x^n\partial_k x^p B_{mnp}
).\label{sc}
\ee
where $M^3$ is $T^2\times { R}$, and spacetime is $M^7\times K3$.
In terms of compact indices, $a,b,...$
 and spacetime indices, $m,n,...$,  action (\ref{sc}) becomes
\ba
S'_M&=&S'_{st}+S'_{KK}+S'_{mod}\non\\
S'_{st}&=&\int d^3\sigma
\sqrt{-g_{mn}\partial_ix^m\partial_jx^n}+
\frac{1}{6} B_{mnp}\epsilon^{ijk}\partial_ix^m\partial_jx^n
\partial_kx^p\non\\
S'_{KK}&=&\frac{1}{6}\int d^3\sigma
\epsilon^{ijk}\partial_i x^{a}\partial_j x^{b} \partial_k x^{m} B_{abm}
\non\\
S'_{mod}&=&
\int d^3\sigma
 \sqrt{-g_{ab}\partial_ix^a\partial_jx^b}+
\frac{1}{6}
\epsilon^{ijk}\partial_i x^{a}\partial_j x^{b} \partial_k x^{c} B_{abc}
\label{IIA}
\ea

The term $S'_{KK}$, has only one relevant term because K3 surfaces have no
one-cycles.
The three-form potential in $S'_{KK}$, in 
analogy with the two-form potential of the heterotic string, generates
KK modes.  The three-form potential that appears in $S_{KK}'$ of
action (\ref{IIA}) 
can be expanded in  terms of the cocycles of K3.
For the 22 two-cocycles of K3, we may decompose $B$ in an 
analogous manner as used in \cite{RW} for the two-form potential
\be
B_{abm}=b_{ab}^{I}(x^a)C^{I}_m(x^r)\label{2c}
\ee
where $I=1,...,22$ labels the two-cycles of $K3$.
Inserting (\ref{2c})
into $S_{KK}'$ yields
\be
\int_{M^3}\epsilon^{ijk}\partial_i x^m\partial_j x^b\partial_k x^a 
b_{ab}^{I}(x^c)C^{I}_m(x^r)
.\label{4c}
\ee
We now use reparametrization invariance to set 
\be
\rho=x^{11},
\ee
where $\rho$ is a worldvolume coordinate, 
and perform a dimensional reduction  of (\ref{4c}).  However, to do so,
we first
review some facts about membrane/string duality of the low energy
theory in D=7 \cite{T,HT}.  

The kinetic terms for the gauge fields in D=7 supergravity read
\be
\int_{M^7}\sqrt{-g^{(7)}}\ a_{IJ}\ F^I_{mn}F^{Jmn}
\ee
and can be obtained after a split of the four-form field strength, $H=dB$, 
of the 11-dimensional supergravity action
\be
H_{abmn}=b^I_{ab}F^I_{mn},
\ee
from the term
\be
\int_{M^{11}}\sqrt{-g^{(11)}}H^2=
\int_{M^7}\sqrt{-g^{(7)}}F^I_{mn}F^{Jmn}\int_{K3}
\sqrt{-g^{(K3)}}b^I_{ab}b^{Jab}.\label{III}
\ee
Membrane/string duality in D=7 requires the existence of a point in the
moduli space of K3  where all the 
22 gauge fields are enhanced\footnote{This also occurs at a point
in the moduli space of the heterotic string on $T^3$.}.
An enhancement of a U(1)
gauge field occurs, as argued in 
\cite{HT}, when a 2-cycle vanishes and two solitonic membranes charged
with respect to a U(1) gauge field  become massless.  
What this means is that even
though a 2-cycle vanishes, the 2-form dual to the vanishing
2-cycle is still well defined.
Otherwise, we see from (\ref{III}), that 
the U(1) symmetry would not be present at
the point in the moduli space since some $b^I_{ab}$ would not 
be defined.  
Therefore, in enhancing
simultaneously the 22 gauge symmetries
we must require the
22 2-cycles of K3 to vanish and the
existence of  a limiting procedure
in which the 22 elements, $b^I$,
of $H^2(K3)$ are well defined at that point in the
moduli where such an enhancement of symmetry takes place.  
Otherwise membrane/string
duality would not hold.  
The 22 two-cocycles, $b^J$ are defined only over their
dual two-cycles, $\gamma_J$, through the relation 
\be
\int_{\gamma_I} b^J=\delta_{IJ}.
\ee
Then, we expect
that the 22 $b^I$'s will be coordinate independent 
when the 22 $\gamma_I$'s vanish to a point.
Therefore, at the point in the moduli space when the 22
two-cycles vanish it holds that
\ba
\partial_{x^{11}}b_{ab}^I&=&0\non\\
\partial_{x^{11}}g_{ab}^I&=&0,\label{tri}
\ea
since the metric, $g_{ab}$,  on K3 is also defined in terms of elments of 
$H^2(K3)$.
At the point in the  moduli space where (\ref{tri}) holds, we
can also require that 
\be
\partial_{\rho}x^M=0\;\;M\ne11.
\ee
This conditions are sufficient to perform a dimensional reduction which
yields
\be
\int_{M^2}\epsilon^{ij}\partial_i x^m\partial_j x^b
b_{11b}^{I}C^{I}_m.
\label{5c}
\ee
Thus, identifying $C^{I}$ with $A^{I}$ and $\partial_j x^bb_{11b}^{I}$
with $\partial_jx^{I}$, we are able to match the gauge sector of the close
membrane with the gauge sector of the open membrane.  This means that the
map
\ba
b^I_{a 11}\partial_j x^a&\to &\partial_jx^I\non\\
C^I_m&\to& A^I_m,
\label{map}
\ea 
which we will refer to as S-duality map,
takes action (\ref{5c}) to the form
\be
\int_{M^2}\epsilon^{ij}\partial_i x^I
A^{I}_m.
\label{6c}
\ee
which is equivalent to the term $S_{KK}$ in (\ref{hete}).  
The map (\ref{map}) does not act on $x^{11}$.
This map acts on the
induced metric on the worldvolume.  Thus, the term in $S_{mod}'$ in 
(\ref{IIA})  yields after a double  dimensional reduction of $x^{11}$ 
\be
\int d^2\sigma
 \oh (g_{IJ}\eta^{ij} + b_{IJ}\epsilon^{ij})\partial_ix^J\partial_jx^I
\label{smod}
\ee
where
\ba
g_{IJ}&=&g_{ab}b^{11a}_Ib^{11b}_J,\non\\
b_{IJ}&=&B_{ab11}b^{11a}_Ib^{11b}_J.
\ea
Action (\ref{smod}) is equivalent to $S_{mod}$ in eq. (\ref{hete}).
Thus, the S-duality
map which takes $S_{KK}$ to $S_{KK}'$ also takes $S_{mod}$ to dimensionally
reduced $S_{mod}'$.

In order to
arrive to this matching of the gauge sectors of the close
and open membrane, it is necessary to 
generate the gauge fields of the closed membrane before dimensionally
reducing the theory, as opposed to the gauge fields of the open membrane
which are always generated within the two-dimensional  theory.
This explains the origin of strong-weak duality in string theory.  
The strong coupling limit of of type IIA string is 11-dimensional
supergravity which is believed to arise as the long wavelength limit
of supermembrane theory.  Thus, gauge fields present in the three dimensional
theory will be strongly interacting, and will continue to be strongly
interacting after dimensional reduction to a two-dimensional theory.
On the other hand, the open membrane has its gauge fields appearing 
in two dimensional  theories, which are therefore weakly interacting.

We must now consider the spacetime part of the action for the close
membrane (\ref{s}).  The term 
\be
\int_{M^3} \sqrt{-g_{mn}\partial_i x^m\partial_j x^n},
\ee
where ${ij}$ label coordinates on $M^3$,  is independent
of $x^{11}$ and therefore independent of $\rho$.  It 
can be dimensionally reduced using the procedure of \cite{D}  which yields 
\be
\int_{M^2} \sqrt{-g_{mn}\partial_i x^m\partial_j x^n }
\ee
where ${ij}$ now label coordinates on $M^2$.  This term coincides with
the first term in  $S_{st}$ in (\ref{hete}).

The term 
\be
\int_{M^3}\epsilon^{ijk}\partial_i x^m\partial_j x^n\partial_k x^p B_{pnm}.
\label{c3}
\ee
can be mapped to a term  \cite{RW}
\be
\int_W d\Sigma^{mnpq}H_{mnpq}\label{c5}
\ee
where $H=dB$ and $W$ is an element of $H_4(M^7)$.  
This term is topological, and duality of the seven 
dimensional space means that $H^3(M^7)=H^4(M^7)$.  Therefore (\ref{c5})
can be written as 
\be
\int_{*W} d\Sigma^{mnp}H_{mnp}\label{c6}
\ee
where $*W$ is the Hodge dual of $W$.
which  again can be mapped to a term \cite{RW}
\be
\int_{M^2}\epsilon^{ij}\partial_i x^m\partial_j x^n b_{nm}.
\label{c7}
\ee
Thus, the b-term in the space time
string action is a direct consequence of the duality
of the seven dimensional duality between 3- and 4-forms, just as it
is the case for the low energy actions \cite{T}.  We then learn that the
term (\ref{c7}) is equivalent to the second term in $S_{st}$ in (\ref{hete}).

Thus, the dimensional reduction of $S'_{st}$ in (\ref{IIA}) yields the
term $S_{st}$ in (\ref{hete}), and therefore, we have succeeded in mapping
the close membrane action on K3  to the open membrane action on 
$T^3\times S^1/Z^2$.
This
shows that as far as the bosonic massless spectrum is concerned,
the open and close membrane actions are dual to each other when one
is compactified on $T^3\times S^1/Z_2$ and 
the other on K3.  Their bosonic massless
spectra, as we have shown are the same because of the duality of 
seven dimensional spaces, and because the 3-form potential of the close
membrane yields the same number of gauge fields and the
same moduli  as the open membrane compactified on $T^3\times S^1/Z_2$.
It remains to show that both theories have the same massless spectra.  
This will be the case because both theories have the same spacetime
supersymmetry \cite{T}.   That is, the S-duality map must also
map the fermions which are sections of the tangent bundle of one theory
to fermions which are sections of a different tangent bundle, but which
preserve spacetime supersymmetry.

\vspace{1cm}

{\bf Acknowledgements}

I have benefited from discussions with B. Campbell and J.D. Lewis.

\pagebreak

\end{document}